\documentclass[preprin,showpacs,preprintnumbers,eqsecnum,amsmath,amssymb,nofootinbib]{revtex4}
\usepackage{graphics,epsfig}
\usepackage{graphicx}
\usepackage{dcolumn}
\usepackage{bm}

\begin{document}
\baselineskip=0.8 cm
\title{Exact Kerr-like solution and its shadow in a gravity model with spontaneous Lorentz symmetry breaking}

\author{Chikun Ding$^{1,2}$}\email{Chikun_Ding@huhst.edu.cn; dingchikun@163.com}
\author{ Changqing Liu$^1$}\email{lcqliu2562@163.com}
\author{R. Casana$^3$}\email{rodolfo.casana@gmail.com}
\author{ A. Cavalcante$^3$}\email{andre_cavs@hotmail.com}
  \affiliation{$^1$Department of Physics, Hunan University of Humanities, Science and Technology, Loudi, Hunan
417000, P. R. China\\
$^2$Key Laboratory of Low Dimensional
Quantum Structures and Quantum Control of Ministry of Education,
and Synergetic Innovation Center for Quantum Effects and Applications,
Hunan Normal University, Changsha, Hunan 410081, P. R. China\\
$^3$Departamento de F\'{\i}sica, Universidade Federal do Maranh\~{a}o, 65080-805 S\~{a}o Lu\'{\i}s, Maranh\~{a}o, Brazil}

\vspace*{0.2cm}
\begin{abstract}
\baselineskip=0.6 cm
\begin{center}
{\bf Abstract}
\end{center}

We obtain an exact Kerr-like black hole solution by solving the corresponding gravitational field equations in Einstein-bumblebee gravity model where Lorentz symmetry is spontaneously broken once a vector field acquires a vacuum expectation value. Results are presented for the purely radial Lorentz symmetry breaking. In order to study the effects of this breaking, we consider the black hole shadow and find that the radial of the unstable spherical orbit on the equatorial plane $r_c$ decreases with the Lorentz breaking constant $\ell>0$, and increases with $\ell<0$. These shifts are similar to those of Einstein-aether black hole. The effect of the LV parameter on the black hole shadow is that it accelerates the appearance of shadow distortion, and could be detected by the new generation of gravitational antennas.

\end{abstract}

 \maketitle

\vspace*{0.2cm}
\section{Introduction}

After the first discovery of gravitational wave (GW) on September 14, 2015 (GW150914) \cite{abbott},  Laser Interferometer
Gravitational wave Observatory (LIGO) has detected GW for several times. It provides a direct confirmation for the existence of a black hole and, confirms that black hole mergers are common in the universe, and will be observed in large numbers in the near future. On April 10, 2019, the Event Horizon Telescope (EHT) Collaboration announced their first shadow image of a supermassive black hole at the center of a neighboring elliptical M87 galaxy \cite{eht1}. With these two successive breaking discoveries, one can now understand the fundamental nature of spacetime really  through experiments.

For the nature of spacetime, there is a most important principle: Lorentz invariance(LI), which is a pillar of general relativity (GR) and the standard model(SM) of particle physics which are both successful field theories describing universe. The former describes gravitation at the classical level, and the latter depicts particles and other three fundamental interactions at the quantum level. However, LI should not be an exact symmetry at all energies  \cite{mattingly}, particularly when one considering quantum gravity effect, it should not be applicable. Though both GR and SM based on LI and the background of spacetime, they handle their entities in profoundly different manners. GR is a classical field theory in curved spacetime that neglects all quantum properties of particles; SM is a quantum field theory in flat spacetime that neglects all gravitational effects of particles. For collisions of particles of $10^{30}$ eV energy (energy higher than Planck scale), the gravitational interactions predicted by GR are very strong and gravity should not be negligible\cite{camelia}. So in this very high energy scale,  one have to consider merging SM with GR in a single unified theory, known as ``quantum gravity", which remains a challenging task. Lorentz symmetry is a continuous spacetime symmetry and cannot exist in a discrete spacetime. Therefore quantization of spacetime at energies beyond the Planck energy, Lorentz symmetry is invalid and one should reconsider giving up LI.

 Thus, the study of Lorentz violation (LV) is a valuable tool to probe the foundations of modern physics. These studies include LV in the neutrino sector \cite{dai}, the standard-model extension (SME) \cite{colladay}, LV in the non-gravity sector \cite{coleman}, and LV effect on the formation of atmospheric showers \cite{rubtsov}.

 The SME is an effective field theory describing the SM coupled to GR, allowing for dynamical curvature modes, and includes additional terms containing information about the LV occurring at the Plank scale \cite{kostelecky2004}. The LV terms in the SME take the form of Lorentz-violating operators coupled to coefficients with Lorentz indices. The presence of LV in a local Lorentz frame is signaled by a nonzero vacuum value for one or more quantities carrying local Lorentz indices. An explicit theory is the ``bumblebee" model \footnote{To the inspiration for this name, see Ref. \cite{bluhm}. }, where the LV arises from the dynamics of a single vector or axial-vector field $B_\mu$, known as the bumblebee field. It is a subset of Einstein-aether theory and ruled by a potential exhibiting a minimum rolls to its vacuum expectation value. Bumblebee gravity was first used by Kostelecky and Samuel in 1989 \cite{dickinson,kostelecky1989} as a simple model for spontaneous Lorentz violating.

Seeking for black hole solutions are very important works in any theory of gravity, because black holes provide into the quantum gravity realm. In 2018, R. Casana {\it et al} found an exact Schwarzschild-like solution in this bumblebee gravity model and investigated its some classical tests \cite{casana}. Then Rong-Jia Yang {\it el al} study the accretion onto this black hole \cite{yang} and find the LV parameter $\ell$ will slow down the mass
accretion rate. However, rotating black hole solutions are the most relevant subcases for astrophysics. These solutions may be also provide exterior metric for rotating stars. So in the present paper, we try to give an exact Kerr-like solution through solving Einstein-bumblebee equations.

We then study black hole shadow and obtain some deviations from GR and some LV gravity theories. The rest of the paper is organized as follows. In Sec. II we provide the background for the Einstein-bumblebee theory studied in this paper. In Sec. III, we derive the Kerr like solution by solving the gravitational field equations. In Sec. IV, we study its black hole shadow and find some effects of the Lorentz breaking constant $\ell$. Sec. V is devoted to a summary.

\section{Einstein-bumblebee theory}

In the bumblebee gravity theory, the bumblebee vector field $B_{\mu}$ acquires a nonzero vacuum expectation value, under a suitable potential, inducing a spontaneous Lorentz symmetry breaking in the gravitational sector. It is
 described by the  action,
\begin{eqnarray}
\mathcal{S}=
\int d^4x\sqrt{-g}\Big[\frac{1}{16\pi G_N}(\mathcal{R}+\varrho B^{\mu}B^{\nu}\mathcal{R}_{\mu\nu})-\frac{1}{4}B^{\mu\nu}B_{\mu\nu}
-V(B^{\mu})\Big], \label{action}
\end{eqnarray}
where $\varrho$ \footnote{If $\varrho=0$, it is the original KS (Kosteleck\'{y} and Samuel) bumblebee models \cite{kostelecky1989}.} is a real coupling constant (with mass dimension $-1$) which controls the non-minimal gravity interaction to bumblebee field $B_\mu$ (with the mass dimension 1). The bumblebee field strength is defined by
\begin{eqnarray}
B_{\mu\nu}=\partial_{\mu}B_{\nu}-\partial_{\nu}B_{\mu}.
\end{eqnarray}
Lorentz and/or $CPT$ (charge, parity and time) violation is triggered by the potential $V(B^{\mu})$, whose functional form is chosen as
\begin{eqnarray}
V=V(B_{\mu}B^{\mu}\pm b^2),
\end{eqnarray}
in which $b^2$ is a real positive constant. It provides a nonvanishing vacuum expectation value (VEV) for bumblebee field $B_{\mu}$. This potential is supposed to have a minimum at $B^{\mu}B_{\mu}\pm b^2=0$ and $V'(b_{\mu}b^{\mu})=0$ to ensure the breaking of the $U(1)$ symmetry, where the field $B_{\mu}$ acquires a nonzero VEV, $\langle B^{\mu}\rangle= b^{\mu}$. The vector $b^{\mu}$ is a function of the spacetime coordinates and has constant magnitude $b_{\mu}b^{\mu}=\mp b^2$, where $\pm$ signs mean that $b^{\mu}$ is timelike or spacelike, respectively.

The action (\ref{action}) yields the gravitational field equation in vacuum
\begin{eqnarray}\label{einstein0}
\mathcal{R}_{\mu\nu}-\frac{1}{2}g_{\mu\nu}\mathcal{R}=\kappa T_{\mu\nu}^B,
\end{eqnarray}
where $\kappa=8\pi G_N$ and the bumblebee energy momentum tensor $T_{\mu\nu}^B$ is \footnote{Its first term should be plus sign as compared to that in Refs \cite{kostelecky2004,casana}.}
\begin{eqnarray}\label{momentum}
&&T_{\mu\nu}^B=B_{\mu\alpha}B^{\alpha}_{\;\nu}-\frac{1}{4}g_{\mu\nu} B^{\alpha\beta}B_{\alpha\beta}- g_{\mu\nu}V+
2B_{\mu}B_{\nu}V'\nonumber\\
&&+\frac{\varrho}{\kappa}\Big[\frac{1}{2}g_{\mu\nu}B^{\alpha}B^{\beta}R_{\alpha\beta}
-B_{\mu}B^{\alpha}R_{\alpha\nu}-B_{\nu}B^{\alpha}R_{\alpha\mu}\nonumber\\
&&+\frac{1}{2}\nabla_{\alpha}\nabla_{\mu}(B^{\alpha}B_{\nu})
+\frac{1}{2}\nabla_{\alpha}\nabla_{\nu}(B^{\alpha}B_{\mu})
-\frac{1}{2}\nabla^2(B^{\mu}B_{\nu})-\frac{1}{2}
g_{\mu\nu}\nabla_{\alpha}\nabla_{\beta}(B^{\alpha}B^{\beta})\Big].
\end{eqnarray}
The prime denotes differentiation with respect to the argument,
\begin{eqnarray}
V'=\frac{\partial V(x)}{\partial x}\Big|_{x=B^{\mu}B_{\mu}\pm b^2}.
\end{eqnarray}
Using the trace of Eq. (\ref{einstein0}), we obtain the trace-reversed version
\begin{eqnarray}\label{einstein}
\mathcal{R}_{\mu\nu}=\kappa T_{\mu\nu}^B+2\kappa g_{\mu\nu}V
-\kappa g_{\mu\nu} B^{\alpha}B_{\alpha}V'+\frac{\varrho}{4}g_{\mu\nu}\nabla^2(B^{\alpha}B_{\alpha})
+\frac{\varrho}{2}g_{\mu\nu}\nabla_{\alpha}\nabla_{\beta}(B^{\alpha}B^{\beta}).
\end{eqnarray}

The equation of motion for the bumblebee field is
\begin{eqnarray}
\nabla ^{\mu}B_{\mu\nu}=2V'B_\nu-\frac{\varrho}{\kappa}B^{\mu}R_{\mu\nu}.
\end{eqnarray}

In the remainder of this manuscript, we assume that the bumblebee field is frozen at its VEV, i.e., it is fixed to be
\begin{eqnarray}
B_\mu=b_\mu,
\end{eqnarray}
then the particular form of the potential driving its dynamics is irrelevant.
And consequently, we have $V=0,\;V'=0$. Then the first both terms in Eq. (\ref{momentum}) are like those of the electromagnetic field, the only difference are the coupling terms to Ricci tensor. Under this condition,  Eq. (\ref{einstein}) leads to gravitational field equations
\begin{eqnarray}\label{bar}
\bar R_{\mu\nu}=0,
\end{eqnarray}
with
\begin{eqnarray}\label{barb}
&&\bar R_{\mu\nu}=\mathcal{R}_{\mu\nu}-\kappa b_{\mu\alpha}b^{\alpha}_{\;\nu}+\frac{\kappa}{4}g_{\mu\nu} b^{\alpha\beta}b_{\alpha\beta}+\varrho b_{\mu}b^{\alpha}\mathcal{R}_{\alpha\nu}
+\varrho b_{\nu}b^{\alpha}\mathcal{R}_{\alpha\mu}
-\frac{\varrho}{2}g_{\mu\nu}b^{\alpha}b^{\beta}\mathcal{R}_{\alpha\beta}+\bar B_{\mu\nu},\nonumber\\
&&\bar B_{\mu\nu}=-\frac{\varrho}{2}\Big[
\nabla_{\alpha}\nabla_{\mu}(b^{\alpha}b_{\nu})
+\nabla_{\alpha}\nabla_{\nu}(b^{\alpha}b_{\mu})
-\nabla^2(b_{\mu}b_{\nu})\Big].
\end{eqnarray}
In the next section, we find the rotating black hole solution by using an elementary method in this Einstein-bumblebee model.

\section{Exact Kerr-like solution in Einstein-bumblebee model}
In this section, we will give the exact Kerr-like solution through solving Einstein-bumblebee equations.

Rotating black hole solutions are the most relevant subcases for astrophysics. These solutions may be also provide exterior metric for rotating stars. However, the generation of such exact rotating solution to Einstein's field equations is very difficult due to the highly non-linear differential equations. Schwarzschild black hole solution was published in 1916 soon after GR was discovered \cite{sch}. But 47 years later, in 1963, the rotating black hole solution was found by Kerr \cite{kerr}. So it is frequently alleged that the Kerr metric cannot be derived by elementary methods (by inference from \cite{misner}, p.877). But in 1982, Klotz used an elementary method to reproduce Kerr solution \cite{klotz}. Then it is used to derive Kerr-Newman \cite{gui} and five dimensional Myers-Perry metric \cite{peng}. In this method,
the  radiating stationary axially symmetric black hole metric  have the general form \cite{klotz}
\begin{eqnarray}\label{me}
&&ds^2=-\gamma(\zeta,\theta)d\tau^2+a[p(\zeta)-q(\theta)]\Big(d\zeta^2+d\theta^2+
\frac{q}{a}d\phi^2\Big)-2q(\theta)d\tau d\phi,
\end{eqnarray}
where $a$ is a constant inserted for dimensional reasons.
The time $t$ is given by
\begin{eqnarray}\label{}
d \tau=dt-q d\phi,
\end{eqnarray}
then Eq. (\ref{me}) becomes
\begin{eqnarray}\label{metric}
&&ds^2=-\gamma(\zeta,\theta)dt^2+a[p(\zeta)-q(\theta)](d\zeta^2+d\theta^2)\nonumber\\
&&+\{[1-\gamma(\zeta,\theta)]q^2(\theta)+p(\zeta)q(\theta)\}d\phi^2
-2q(\theta)[1-\gamma(\zeta,\theta)]dtd\phi.
\end{eqnarray}
We will use this metric ansatz to set up gravitational field equations.

In this study, we focus on that the bumblebee field acquiring a purely radial vacuum energy expectation since the spacetime curvature has a strong radial variation when compared with very slow temporal changes. So the bumblebee field is spacelike and assumed to be
\begin{eqnarray}
b_\mu=(0,b(\zeta,\theta),0,0).
\end{eqnarray}
By using the condition $b_\mu b^\mu=$constant, the explicit form of $b_\mu$ is
\begin{eqnarray}\label{bu}
b_\mu=(0,b_0\sqrt{a(p-q)}\;,0,0),
\end{eqnarray}
where $b_0$ is a constant. Note that it is different from the electromagnetic field $A_{\mu}$ \cite{gui}.
The bumblebee field strength is
\begin{eqnarray}
b_{\mu\nu}=\partial_{\mu}b_{\nu}-\partial_{\nu}b_{\mu}.
\end{eqnarray}
Its nonzero components are $b_{\zeta\theta}=-b_{\theta \zeta}=ab_0q'/2\sqrt{a(p-q)}$, where the prime denotes to derivative to its argument. The other nonzero components of the quantity $b^\alpha_\mu b_{\nu\alpha}$ are $b^\alpha_\zeta b_{\zeta\alpha}=b^\alpha_\theta b_{\theta\alpha}=b_0^2q'^2/4(p-q)^2$. And the quantity $b^{\alpha\beta} b_{\alpha\beta}=b_0^2q'^2/2a(p-q)^3$.

 For the metric (\ref{metric}), the nonzero components of Ricci tensor are $\mathcal{R}_{tt},\mathcal{R}_{t\phi},\mathcal{R}_{\zeta\zeta},
 \mathcal{R}_{\zeta\theta},\mathcal{R}_{\theta\theta},\mathcal{R}_{\phi\phi}$, shown in the appendix.
 Here we find that $\bar B_{\zeta\theta}=0$ and, give some of gravitational field equations as following
\begin{eqnarray}
&&\bar R_{\zeta\theta}=(1+\ell) \mathcal{R}_{\zeta\theta}\\
&&\bar R_{tt}=\mathcal{R}_{tt}+g_{tt}\Big(\frac{\kappa}{4}b^{\alpha\beta} b_{\alpha\beta}-\frac{\varrho}{2}b^\zeta b^\zeta \mathcal{R}_{\zeta\zeta}\Big)+\bar B_{tt},\label{tt}\\
&&\bar R_{t\phi}=\mathcal{R}_{t\phi}+g_{t\phi}\Big(\frac{\kappa}{4}b^{\alpha\beta} b_{\alpha\beta}-\frac{\varrho}{2}b^\zeta b^\zeta \mathcal{R}_{\zeta\zeta}\Big)+\bar B_{t\phi},\label{tp}
\end{eqnarray}
where $\ell=\varrho b^2_0$.
The quantities $\mathcal{R}_{\zeta\theta},\bar B_{tt},\bar B_{t\phi}$ are as
\begin{eqnarray}
&&\mathcal{R}_{\zeta\theta}=-\frac{\bar\Delta_{12}}{2\bar\Delta}
+\frac{\Delta_{2}[(p-q)\bar\Delta_1+2\bar\Delta p_1]}{4(p-q)\bar\Delta^2},\label{rth}\\
&&\bar B_{tt}=\ell\Big[\frac{\gamma_{11}}{2a(p-q)}
+\frac{\gamma}{4a(p-q)\bar\Delta}p_1\gamma_1-\frac{1}{4a\bar\Delta}\gamma_1^2\Big],\\
&&\bar B_{t\phi}=\ell\Big[-\frac{q\gamma_{11}}{2a(p-q)}
+\frac{q(2-\gamma)}{4a(p-q)\Delta}p_1\gamma_1+\frac{q}{4a\bar\Delta}\gamma_1^2\Big],
\end{eqnarray}
where $\bar\Delta=q+\gamma (p-q)$, and the derivatives with respect to $\zeta$ and $\theta$ are denoted by the suffixes 1 and 2, respectively.

$\bar R_{\zeta\theta}=0$ showing that $\mathcal{R}_{\zeta\theta}$ is zero, then from Eq. (\ref{rth}), we can assume that $\bar\Delta_2=0$ or
\begin{eqnarray}
\gamma_2=-\frac{(1-\gamma)q_2}{p-q}.
\end{eqnarray}
Then the function $\gamma$ can be given by
\begin{eqnarray}
\gamma=1-\frac{2h(\zeta)}{p(\zeta)-q(\theta)}.
\end{eqnarray}
The condition $\bar\Delta_2=0$ enables us to introduce a new independent variable,
\begin{eqnarray}
\sigma=\int \sqrt{\bar\Delta}d\zeta,
\end{eqnarray}
where $\bar\Delta=p-2h$. So, derivatives with respect to $\zeta$ become
\begin{eqnarray}
p_1=\frac{dp}{d\zeta}=\frac{d\sigma}{d\zeta}\frac{dp}{d\sigma}=\sqrt{\bar\Delta}
\frac{dp}{d\sigma},\;p_{11}=\frac{d^2p}{d\zeta^2}=\bar\Delta\frac{d^2p}{d\sigma^2}+\frac{1}{2}
\big(\frac{dp}{d\sigma}\big)^2-\frac{dh}{d\sigma}\frac{dp}{d\sigma}.
\end{eqnarray}
From the Eqs. (\ref{bar}), (\ref{tt}) and (\ref{tp}), we can find the combination that
\begin{eqnarray}
g_{t\phi}\bar R_{tt}-g_{tt}\bar R_{t\phi}=0.
\end{eqnarray}
This combination can be reduced to
\begin{eqnarray}\label{pq}
&&p\Big[4(1+\ell)\frac{\dot{h}\dot{p}}{h}q^2-2qq_{22}+q_2^2+2(1+\ell)\ddot{p}q^2\Big]
\nonumber\\&&
~~~-4(1+\ell)\dot{p}^2q^2-2(p-q)^2q^2\ddot{h}(1+\frac{\ell}{h})\nonumber\\&&
-q\Big[4(1+\ell)\frac{\dot{h}\dot{p}}{h}q^2-2qq_{22}+5q_2^2+2(1+\ell)\ddot{p}q^2\Big]=0,
\end{eqnarray}
where dots denote derivatives with respect to $\sigma$. Note that $p$ and $h$ are functions of $\sigma$ only, and $q$ is a $\theta$ function, so we must have
\begin{eqnarray}\label{fpk}
\frac{\dot{h}\dot{p}}{h}=k,\;\dot{p}^2=cp+n,\;\ddot{h}=0,
\end{eqnarray}
where $k,c,n$ are some constants. Then $\ddot{p}=k=c/2$ and Eq. (\ref{pq}) can be reduced to
\begin{eqnarray}\label{}
&&4(1+\ell)(k-c)q^2-2qq_{22}+q_2^2+(1+\ell)cq^2=0,
\nonumber\\&&
4k(1+\ell)q^2-2qq_{22}+5q_2^2+(1+\ell)cq^2+4(1+\ell)nq=0.
\end{eqnarray}
They both give
\begin{eqnarray}\label{}
q^2_2=-(1+\ell)(cq^2+nq).
\end{eqnarray}
We can obtain that
\begin{eqnarray}\label{}
q=-\frac{n}{c}\sin^2[\sqrt{(1+\ell)c}\theta/2].
\end{eqnarray}
By setting the constants $c=4/(1+\ell)$ and $n=-4a$, it becomes
\begin{eqnarray}\label{sigma1}
q=(1+\ell)a\sin^2\theta.
\end{eqnarray}
From the conditions (\ref{fpk}), we find that
\begin{eqnarray}\label{sigma2}
p=\frac{\sigma^2}{1+\ell}+a(1+\ell),\;h=c'\sigma,
\;\gamma=1-\frac{2(1+\ell)c'\sigma}{\sigma^2+a(1+\ell^2\cos^2\theta)},
\end{eqnarray}
where $c'$ is a constant. After choosing $\sigma=\sqrt{(\ell+1)/a}r$, $c'=M/\sqrt{(\ell+1)a}$ and $\phi=\varphi/\sqrt{1+\ell}$ for Boyer-Lindquist coordinates, we can get that
\begin{eqnarray}\label{}
p=\frac{r^2}{a}+a(\ell+1),\;h=\frac{Mr}{a},\;\gamma=1-\frac{2Mr}{\rho^2},
\end{eqnarray}
where $\rho^2=r^2+(1+\ell)a^2\cos^2\theta$.
Lastly, substituting these quantities into Eqs. (\ref{metric}) and (\ref{bu}), we can get the bumblebee field $b_\mu=(0,b_0\rho,0,0)$, and the
 rotating metric in the bumblebee gravity
\begin{eqnarray}\label{bmetric}
ds^2=- \Big(1-\frac{2Mr}{\rho^2}\Big)dt^2-\frac{4Mra\sqrt{1+\ell}\sin^2\theta}{\rho^2}
dtd\varphi+\frac{\rho^2}{\Delta}dr^2+\rho^2d\theta^2
+\frac{A\sin^2\theta}{\rho^2} d\varphi^2,
\end{eqnarray}
where
\begin{eqnarray}
\Delta=\frac{r^2-2Mr}{1+\ell}+a^2,\;A=\big[r^2+(1+\ell)a^2\big]^2-\Delta(1+\ell)^2 a^2\sin^2\theta.
\end{eqnarray}
If $\ell\rightarrow0$, it recovers the usual Kerr metric.
When $a\rightarrow0$, it becomes
\begin{eqnarray}
ds^2=- \Big(1-\frac{2M}{r}\Big)dt^2+\frac{1+\ell}{1-2M/r}dr^2+r^2d\theta^2
+r^2\sin^2\theta d\varphi^2,
\end{eqnarray}
which is the same as that in Ref. \cite{casana}.
The metric (\ref{bmetric}) represents a purely radial Lorentz-violating  black hole solution with rotating angular momentum $a$. It is singular at $\rho^2=0$ and at $\Delta=0$. The solution of $\rho^2=0$  is a ring shape physical
singularity at the equatorial plane of the center of rotating black hole with radius $a$. Its event horizons and ergosphere  locate at
\begin{eqnarray}
r_{\pm}=M\pm\sqrt{M^2-a^2(1+\ell)},\;r^{ergo}_{\pm}=M\pm\sqrt{M^2-a^2(1+\ell)\cos^2\theta},
\end{eqnarray}
where $\pm$ signs correspond to outer and inner horizon/ergosphere, respectively. It is easy to see that there exists a black hole if and only if
\begin{eqnarray}
|a|\leq \frac{M}{\sqrt{1+\ell}}.
\end{eqnarray}
Its Hawking temperature can be obtained from its surface gravity \cite{wald}
\begin{eqnarray}
T=\frac{\kappa}{2\pi},\;\kappa=-\frac{1}{2}\lim_{r\rightarrow r_+}\sqrt{\frac{-1}{X}}\frac{dX}{dr},\;X\equiv g_{tt}-\frac{g^2_{t\varphi}}{g_{\varphi\varphi}}.
\end{eqnarray}
Inserting corresponding metric components in Eq. (\ref{bmetric}), one get
 \begin{eqnarray}
T=\frac{\sqrt{1+\ell}\Delta'(r_+)}{4\pi [r_+^2+(1+\ell)a^2]}=\frac{r_+-M}{2\pi\sqrt{1+\ell} \big[r_+^2+(1+\ell)a^2\big]}.
\end{eqnarray}

\section{Black hole shadow}
In this section, we study some observational signatures on the Lorentz-violating parameter $\ell$ by analyzing black hole shadow with the metric (\ref{bmetric}), and try to find some deviation from GR and some similarities to other LV black holes.

\begin{figure}[ht]
\begin{center}
\includegraphics[width=7.0cm]{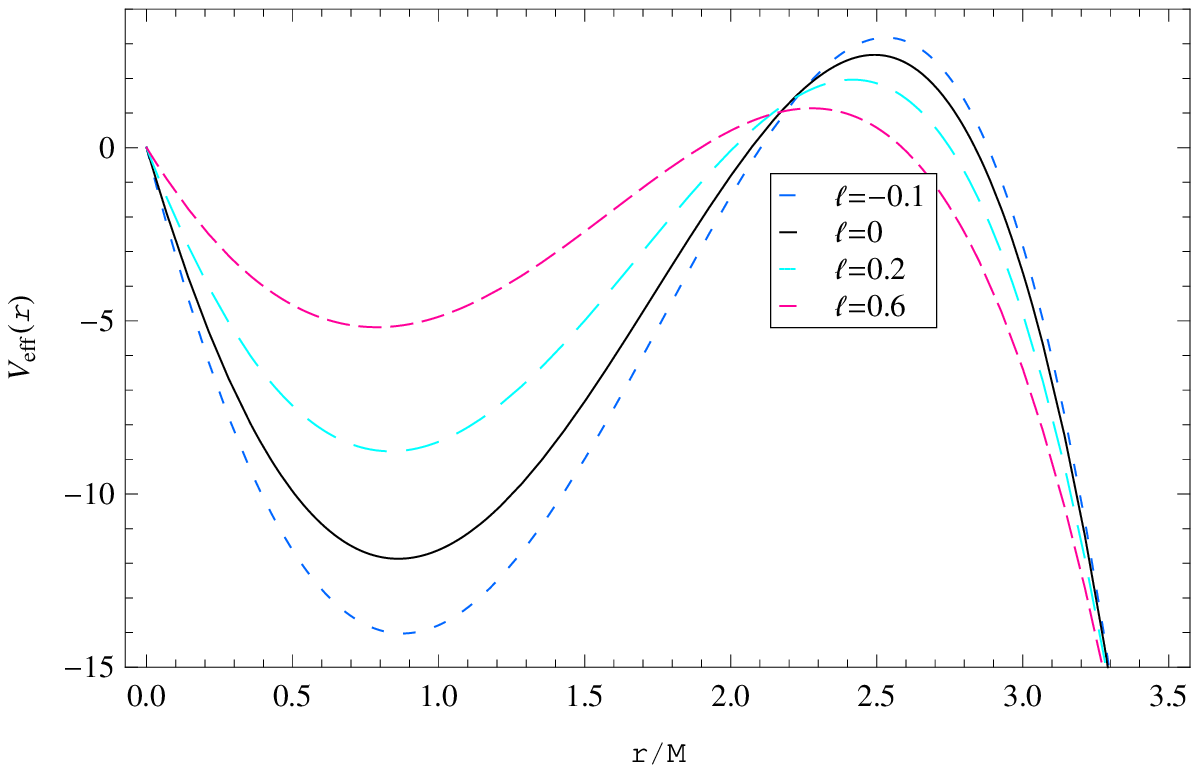}\;\;\includegraphics[width=7.0cm]{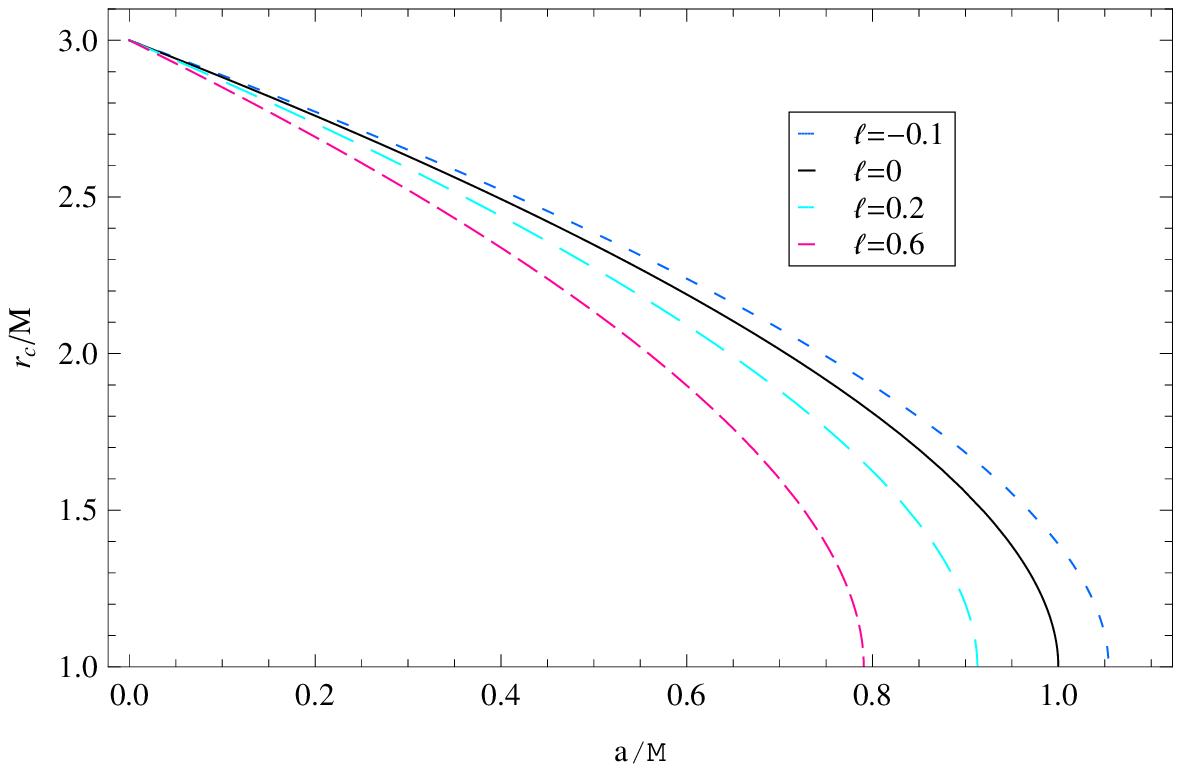}
\caption{The left panel describes the effective potential. The right one shows the radius $r_c$ of the equatorial circular orbit.}\label{fp1}
 \end{center}
 \end{figure}

We introduce two conserved parameters $\xi$ and $\eta$ by
\begin{eqnarray}
\xi=\frac{L_z}{E},\;\eta=\frac{\mathcal{Q}}{E^2},
\end{eqnarray}
where $E,\;L_z$ and $\mathcal{Q}$ are the energy, axial component of the angular momentum and Carter constant, respectively.
Then the null geodesics in the bumblebee rotating black hole spacetime are given by
\begin{eqnarray}\label{geod}
&&\rho^2\frac{dr}{d\lambda}=\pm\sqrt{R},
\;\rho^2\frac{d\theta}{d\lambda}=\pm\sqrt{\Theta},
\nonumber\\
&&(1+\ell)\Delta\rho^2\frac{dt}{d\lambda}
=A-2\sqrt{1+\ell}Mra\xi,\nonumber\\
&&(1+\ell)\Delta\frac{d\phi}{d\lambda}
=2\sqrt{1+\ell}Mra+\frac{\xi}{\sin^2\theta}(\rho^2-2Mr),
\end{eqnarray}
where $\lambda$ is the affine parameter and,
\begin{eqnarray}
R(r)=[X(r)-a\xi]^2-\Delta[\eta+(\xi-\sqrt{1+\ell}a)^2],\;
\Theta(\theta)=\eta+(1+\ell)a^2\cos^2\theta-\xi^2\cot^2\theta,
\end{eqnarray}
with $X(r)=[r^2+(1+\ell)a^2]/\sqrt{1+\ell}$.
The radial motion in Eqs. (\ref{geod}) can be written in the form
\begin{eqnarray}
\Big(\rho^2\frac{dr}{d\lambda}\Big)^2+V_{eff}=0,
\end{eqnarray}
which is similar to the equation of motion of a classical particle. The effective potential is
\begin{eqnarray}
V_{eff}=-\frac{r^4}{1+\ell}+\Big(\frac{\eta+\xi^2}{1+\ell}-a^2\Big)r^2
-2M\Big[\Big(\frac{\xi}{\sqrt{1+\ell}}-a\Big)^2+\frac{\eta}{1+\ell}\Big]r+a^2\eta,
\end{eqnarray}
which has the limit $V_{eff}(0)=0,\;V_{eff}(r\rightarrow\infty)\rightarrow-\infty$. We plot the $V_{eff}$ against $r$ in Fig. \ref{fp1} with $\eta=0,\;a/M=0.5$ and $\xi=\xi_c+0.2$.

Fig. \ref{fp1} shows that the photon starting from infinity will meet a turning point, and then turns back to infinity. When $\xi=\xi_c$, this turning point is an unstable spherical orbit which gives the boundary of the shadow \cite{wei}. Fig. \ref{fp1} also shows that the deviation from GR (Kerr): when LV constant $\ell>0$, the turning point shifts to the left; when $\ell<0$, it shifts to the right. These shifts are similar to those of the Einstein-aether black hole \cite{tao}, which is also a LV black hole.

The unstable spherical orbit on the equatorial plane is given by the following equations
\begin{eqnarray}\label{orbit}
\theta=\frac{\pi}{2},\;R(r)=0,\;\frac{dR}{dr}=0\;,\frac{d^2R}{dr^2}<0\;,\eta=0,
\end{eqnarray}
which give the radius of the unstable orbit as
\begin{eqnarray}
r_c^{\pm}=2M(1+\cos2\theta),\;\theta=\frac{1}{3}\arccos[\mp \sqrt{1+\ell}a/M],
\;\xi_c=6M\cos\theta-\sqrt{1+\ell}a,
\end{eqnarray}
where the upper sign is to direct orbits and the lower sign to retrograde orbits.
We plot the equatorial circular orbit $r_c$ against $a$ in Fig. \ref{fp1}. It shows that the $r_c$ decreases with $\ell>0$, and increases with $\ell<0$, which are similar to those of the noncommutative black hole \cite{wei}.

\begin{figure}[ht]
\begin{center}
\includegraphics[width=7.0cm]{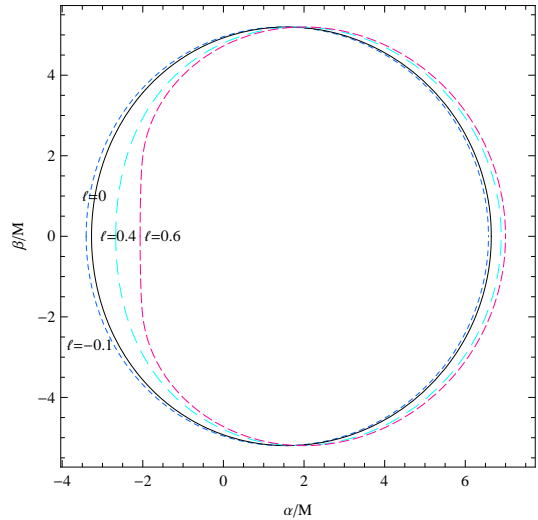}
\caption{The shapes of the shadow with $a/M=0.79,\;\theta=\pi/2$. The black solid line is for Kerr black hole shadow.}\label{fp2}
 \end{center}
 \end{figure}

For more generic orbits $\theta\neq\pi/2$ and $\eta\neq0$, the solution of Eq. (\ref{orbit}), $r=r_s$, gives the $r-$constant orbit, which is also called spherical orbit. And the both conserved parameters of the spherical orbits can be written as
\begin{eqnarray}
\xi_s=\frac{r_s^2(3M-r_s)-(1+\ell)a^2(M+r_s)}{\sqrt{1+\ell}a(r_s-M)},
\;\eta_s=\frac{r_s^3[4(1+\ell)Ma^2-r_s(r_s-3M)^2]}{(1+\ell)a^2(r_s-M)^2}.
\end{eqnarray}
Nextly, the two celestial coordinates, which are used to describe the shape of the shadow that an observers seen in the sky, can be given by
\begin{eqnarray}
\alpha=-\xi_s\csc\theta,\;\beta=\sqrt{\eta_s+a^2\cos^2\theta-\xi_s^2\cot^2\theta}.
\end{eqnarray}
We show the shapes of the shadow in Fig. \ref{fp2}.

The Fig. \ref{fp2} shows that the distortion of the shadow when this LV black hole rotates fast $a/M=0.79$. With the increase of the LV parameter $\ell$, its left endpoint moves to the right obviously, and then the right endpoint moves to the right slightly. As for Kerr black hole, the similarly distortion occurs till $a/M>0.98$. So this is the effect of the LV parameter on the black hole shadow, i.e., accelerating the appearance of shadow distortion. If this LV parameter $\ell$ is not very small, the derivation of black hole shadow from Kerr black hole may be observed in the near future black hole shadow image events.

\section{Summary}

In this paper, we have studied the stationary, axisymmetric, asymptotically flat black hole solutions of Einstein-bumblebee theory in the $4$-dimensional spacetime. In this model, a vector field, termed bumblebee, couples to the spacetime curvature and acquires a vacuum expectation value, which induces Lorentz symmetry spontaneously broken. In the case of purely radial Lorentz symmetry breaking, we have achieved a new exact rotating solution to the gravitational field equations. When angular momentum $a\rightarrow0$, it can recover Schwarzschild like solution \cite{casana}; when LV constant $\ell\rightarrow0$, it can recover Kerr black hole solution. We then give the positions of horizons and  its Hawking temperature.

With this given black hole solution, we can find some LV effects by astronomical observations. In order to obtain these effects of LV constant $\ell$, we study the black hole shadow since the first shadow image of a black hole is released by EHT Collaboration on April 10, 2019 \cite{eht1}. It shows that the deviation from GR (Kerr black hole): when LV constant $\ell>0$, the turning point of the effective potential shifts to the left (or the equatorial circular orbit $r_c$ decreases); when $\ell<0$, it shifts to the right (or the equatorial circular orbit $r_c$ increases). These shifts are similar to those of the Einstein-aether black hole \cite{tao}, which is also a LV black hole. And the effect of the LV parameter on the black hole shadow is that it accelerates  the appearance of shadow distortion. These difference could be detected by the new generation of gravitational antennas.

\begin{acknowledgments} The authors thank the National Natural Science Foundation (NNSFC)
of China (grant No. 11247013), Hunan Provincial Natural Science Foundation of China (grant No. 2015JJ2085), the Scientific Research Fund of the Hunan Provincial Education Department under No. 19A257, CAPES, CNPq, and FAPEMA (Brazilian agencies) for financial support.
\end{acknowledgments}
\appendix\section{Some quantities}
In this appendix, we showed the covariant derivatives with $b^{\alpha }b_{\nu}$ in Eq. (\ref{barb}) and the nonezero components of Ricci tensor for the metric (\ref{metric}).
\begin{eqnarray}
\nabla_{\alpha}\nabla_{\mu}(b^{\alpha}b_{\nu})
&=&\partial_{\alpha}[\nabla_{\mu}(b^{\alpha}b_{\nu})]
+\Gamma^{\alpha}_{\alpha\tau}\nabla_{\mu}(b^{\tau}b_{\nu})
-\Gamma^{\tau}_{\alpha\mu}\nabla_{\tau}(b^{\alpha}b_{\nu})
-\Gamma^{\tau}_{\alpha\nu}\nabla_{\mu}(b^{\alpha}b_{\tau})\nonumber\\
&=&\partial_{\alpha}[\partial_{\mu}(b^{\alpha}b_{\nu})
-\Gamma^{\tau}_{\mu\nu}b^{\alpha}b_{\tau}+\Gamma^{\alpha}_{\mu\tau}b^{\tau}b_{\nu}]
+\Gamma^{\alpha}_{\alpha\tau}[\partial_{\mu}(b^{\tau}b_{\nu})
+\Gamma^{\tau}_{\mu\rho}b^{\rho}b_{\nu}-\Gamma^{\rho}_{\mu\nu}b^{\tau}b_{\rho}]\nonumber\\
&&-\Gamma^{\tau}_{\alpha\mu}[\partial_{\tau}(b^{\alpha}b_{\nu})
+\Gamma^{\alpha}_{\tau j}b^{j}b_{\nu}-\Gamma^{j}_{\tau\nu}b^{\alpha}b_{j}]
-\Gamma^{\tau}_{\alpha\nu}[\partial_{\mu}b^{\alpha}b_{\tau}
+\Gamma^{\alpha}_{\mu k}b^{k}b_{\tau}-\Gamma^{k}_{\mu\tau}b^{\alpha}b_{k}],\\
\nabla^2(b_{\mu}b_{\nu})&=&g^{\alpha\tau}\nabla_{\alpha}\nabla_{\tau}(b_{\mu}b_{\nu}).
\end{eqnarray}
\begin{eqnarray}
\mathcal{R}_{tt}=&&\frac{\gamma_{11}+\gamma_{22}}{2 \Sigma}-\frac{1}{4 \bar\Delta\Sigma}\left[\left(\gamma_{1}^{2}+\gamma_{2}^{2}\right)(p-q)+2 \gamma_{2} q_{2}(1-\gamma)-\gamma p_{1} \gamma_{1}\right]\nonumber \\
&&+\frac{\gamma q_{2}}{4 q \bar\Delta\Sigma}\left[-2 q_{2}(1-\gamma)^{2}-p \gamma_{2}\right],\\
\mathcal{R}_{t\phi}=&& -\frac{q\left(\gamma_{11}+\gamma_{22}\right)}{2 \Sigma}+\frac{q}{4 \bar\Delta\Sigma}\left[p_{1} \gamma_{1}(2-\gamma)+3 \gamma_{2} q_{2}(1-\gamma)+(p-q)\left(\gamma_{1}^{2}+\gamma_{2}^{2}\right)\right] \nonumber\\
&&-\frac{1}{4 \bar\Delta\Sigma}\left\{2 \bar\Delta\left[2 \gamma_{2} q_{2}-(1-\gamma) q_{22}\right]+\left[2 \gamma q_{2}^{2}(1-\gamma)^{2}-\gamma_{2} p q_{2}\right]\right\}-\frac{1}{4 q \bar\Delta\Sigma} \gamma p q_{2}^{2}(1-\gamma),\\
\mathcal{R}_{\zeta\zeta}=&&-\frac{p-q}{2 \bar\Delta}\gamma_{11}+\frac{1}{4 \bar\Delta }(q_2\gamma_2-p_1\gamma_1)+\frac{(p-q)^2}{4\bar\Delta^2}\gamma_1^2+
\left[\frac{5\bar\Delta-3q}{4(p-q)^2}+\frac{\gamma^2}{q} \right]\frac{p_1^2}{\bar\Delta^2}\nonumber \\
&&+\frac{p(q+\bar\Delta)}{4 q \bar\Delta(p-q)^2}q_2^2+\frac{(2\bar\Delta-3q)}{2 \bar\Delta(p-q)}p_{11}+\frac{q_{22}}{2 (p-q)},\\
\mathcal{R}_{\theta\theta}=&&-\frac{p-q}{2 \bar\Delta}\gamma_{22}+\frac{1}{4 \bar\Delta }\left(\frac{2p+3\bar\Delta}{\bar\Delta}q_2\gamma_2-p_1\gamma_1\right)
+\frac{(p-q)^2}{4\bar\Delta^2}\gamma_2^2+\frac{q+\bar\Delta}{4  \bar\Delta(p-q)^2}p_1^2+\frac{(pq+q\bar\Delta-3q\bar\Delta)}{2q \bar\Delta(p-q)}q_{22}-\frac{p_{11}}{2 (p-q)}\nonumber \\
&&+
\left[\frac{2(p-q)}{q}\gamma^3+(p^2-5pq+10q^2)\frac{\gamma^2}{q^2}
+\frac{2p^2-11pq+14q^2}{q(p-q)}\gamma+\frac{2(p^2-3pq+3q^2)}{(p-q)^2} \right]\frac{q_2^2}{4\bar\Delta^2},\\
\mathcal{R}_{\phi\phi}=&& \frac{q^2\left(\gamma_{11}+\gamma_{22}\right)}{2 \Sigma}-\frac{1}{4 \bar\Delta \Sigma}\left[q(p+3q-q\gamma)p_{1} \gamma_{1}+(p^2+pq-4q\bar\Delta-3p\gamma)\gamma_{2} q_{2}-(p-q)q^2\left(\gamma_{1}^{2}+\gamma_{2}^{2}\right)\right] \nonumber\\
&&+\frac{q\gamma}{4 \bar\Delta\Sigma}p_1^2-\frac{1}{2\Sigma}[qp_{11}+(p+2q-2\gamma)q_{22}]+\frac{q_{2}^{2}}{4 q \bar\Delta\Sigma}[p^2\gamma-2q^2 (1-\gamma)^3],
\end{eqnarray}
where $\Sigma=a(p-q)$.

\end{document}